\documentclass[preprint,prl,amsmath,nofootinbib,twoside]{revtex4}
\usepackage{braket}
\usepackage{graphicx}
\usepackage{hyperref}

\begin{document}
\title{Seeking Lorentz Violation from the Higgs}
\date{\today}

\author{Andrew G. Cohen}
\email{cohen@bu.edu}
\author{Gustavo Marques Tavares}
\email{gusmt@bu.edu}
\author{Yiming Xu}
\email{ymxu@bu.edu}
\affiliation{Physics Department,  Boston University\\
  Boston, MA 02215, USA}

\begin{abstract}
  The recently discovered Higgs particle with a mass near $126$ GeV
  presents new opportunities to explore Lorentz
  violation. Ultra-high-energy cosmic rays are one of the most
  sensitive testing grounds for Lorentz symmetry, and can be used to
  seek for and limit departures from Lorentz invariance in the Higgs
  sector.  If the Higgs were to have a super- or sub-luminal maximal
  speed both Higgs and weak interaction physics would be
  modified. Consideration of such modifications allow us to constrain
  the Higgs maximal velocity to agree with that of other Standard
  Model particles to parts in $10^{14}$.
\end{abstract}

\maketitle{}

\section{Introduction and Conclusions}

Departures from Lorentz invariance have been searched for in a wide
variety of circumstances and limited with sometimes astonishing
precision. Some of the best constraints arise from the novel
kinematics associated with departures from a Lorentz invariant
dispersion relation. In a Lorentz invariant world all particles share
a universal maximal speed, usually referred to as the speed of light
and conventionally chosen to be one. The energy and momentum of
particle species $i$ satisfy a Lorentz invariant dispersion relation
$E^{2} = p^{2}+m_{i}^{2}$, and energy and momentum conservation then
forbid many processes; for example, an electron (in vacuum) may not
spontaneously radiate a photon, nor can a photon decay into an
electron-positron pair.

Absent Lorentz invariance, as within a medium, the maximal speeds of
different particle species may differ, $E^{2} = c_{i}^{2} p^{2} +
m_{i}^{2}$. Under these circumstances processes that are otherwise
forbidden may be allowed. For example, in water the maximal speed of
photons is slightly less than it is in vacuum and also less than that
of electrons in the same medium ($c_{\gamma}< c_{e}$). In this case
the emission of photons from superluminal (in the medium) electrons is
both allowed and observed, where it is called Cerenkov radiation. If
the vacuum itself is not Lorentz invariant a similar emission
phenomenon can occur even in the absence of a medium. The observation
of ultra-high-energy cosmic rays from distant sources limits the
energy loss that would accompany this ``vacuum Cerenkov radiation'',
and provides many of the astonishing constraints referred to
above\cite{Coleman:1998ti,Coleman:1997xq}.

The recently discovered Higgs particle provides a new testing ground
for Lorentz violation. Data from the LHC and elsewhere strongly favors
the interpretation of this $126$ GeV mass particle as the radial mode
of a complex $SU(2)$ doublet field, $H$, where the remaining three
fields of the doublet are ``eaten'' via the Higgs mechanism to give
mass to the $W$ and $Z$ bosons. Adopting this interpretation we
consider modifications of the Standard Model Lagrangian involving this
doublet. Imposing rotational, $CPT$ and $CP$ invariance, we find that
there is a unique (subject to field redefinitions) modification of the
Lagrangian with operators of dimension four:
\begin{equation}
  \label{eq:1}
  {\cal L}_{LV} = \delta\  D_{0}H^{\dagger} D^{0} H = \delta \ n\cdot
  DH^{\dagger} n\cdot D H\ .
\end{equation}
where $n^{\mu}$ is a unit timelike four vector and $\delta$ is a
parameter characterizing the size of Lorentz violation.  This addition
to the SM Lagrangian affects the propagation of both the physical
Higgs particle and the $W$ and $Z$ bosons:
\begin{itemize}
\item The maximal speed of the physical Higgs particle is
  $1/\sqrt{1+\delta}$. For positive $\delta$ the Higgs is subluminal;
  for negative $\delta$ it is superluminal.
\item While the propagation of the two transverse polarizations of the
  $W$ and $Z$ bosons are unaffected, the third polarization has the
  same maximal speed as the Higgs particle, $1/\sqrt{1+\delta}$. The
  full vector boson propagator (in unitary gauge) is
  \begin{multline}
    \label{eq:2}
    \frac{-i}{k^{2}-M^{2}}\left\{g^{\mu\nu}-\frac{k^{\mu}k^{\nu}}{M^{2}}
    \right. \\
    \left.
      + \frac{\delta M^{2}}{k^{2}-M^{2}+\delta (k\cdot
        n)^{2}}\left[ n^{\mu} n^{\nu} + (k\cdot
        n)^{2}\frac{k^{\mu}k^{\nu}}{M^{4}} - \frac{(k\cdot
          n)}{M^{2}}\left(k^{\mu} n^{\nu}+k^{\nu} n^{\mu}\right)
      \right] \right\}
  \end{multline}
\end{itemize}

Observations of ultra-high-energy cosmic rays significantly constrain
the size of these modifications. For positive $\delta$ the Higgs
particle's maximal speed is subluminal and stable particles such as
the proton would lose energy via ``vacuum Higgs radiation''. The
observation of high energy protons from distant sources limits such
energy loss.  Higgs emission from a high energy proton is
kinematically allowed when the proton energy is greater than the
threshold energy $E_{0} = M_{H}/\sqrt{\delta}$ (ignoring small
corrections from the proton mass) and the rate of this emission is
readily calculated for small $\delta$. At leading order in $\delta$
the amplitude for a proton to emit a Higgs with energy and momentum $q
= (\epsilon, \mathbf q)$ remains Lorentz invariant, and is then a
function of the Higgs four-momentum squared $q^{2} = \epsilon^{2} -
\vert \mathbf q \vert^{2}$. Using the dispersion relation for the
subluminal Higgs this is $q^{2 }= -\delta( \epsilon^{2}-E_{0}^{2})
\equiv -Q^{2}$. For Higgs energies a few times the threshold energy
$E_{0}$, $q^{2}$ is spacelike and much greater than the QCD
scale. Since $Q^{2}$ is large the proton is likely to fragment as a
result of the emission and the parton model may then be used to
calculate the rate for the inclusive emission process
$\text{\textit{proton}} \to
\text{\textit{Higgs}}+\text{\textit{hadrons}}$.

The rate for inclusive Higgs emission from a proton with momentum $P$
is given by the incoherent sum of the rate for Higgs emission from
each parton species $f$ carrying momentum fraction $x$ of the initial
proton weighted by the parton distribution function $f_{f}(x)$
(ignoring the small effects of the proton mass and QCD corrections)
\begin{equation}
  \label{eq:4}
  \Gamma(P) = \sum_{f}\int^{1}_{0}\! dx f_{f}(x){\hat \Gamma_{f}(x P)}
\end{equation}
The Higgs interacts directly with quarks through the Yukawa couplings,
and indirectly with gluons through the top quark loop. The
contribution to the emission rate from quark $q$ is
\begin{equation}
  \label{eq:5}
  \Gamma =  \frac{M^{2}_{H}}{48\pi E}
  \left(
    \frac{E}{E_{0}} 
  \right)^{2}
  \lambda_{q}^{2 } 
  \int^{1}_{\frac{E_{0}}{E}}\! dx
  \, x f_{q}(x)   (1-\frac{E_{0}}{E}\frac{1}{x})^{2}
  (1+ \frac{E_{0}}{E}\frac{2}{x})
\end{equation}
where $\lambda_{q}$ is the Yukawa coupling for the quark. A similar
expression holds for emission from a gluon. There are significant
uncertainties in the parton distribution functions as $x$ approaches
one, especially for the heavier quarks and gluons. For our
conservative bound we include the contributions from only the up and
down quarks. The contribution of the remaining quarks and the gluons
would only increase the rate.

Ultra-high-energy protons from distant sources have been observed with
energies near $10^{20}$ eV. If this energy is above the threshold
energy for Higgs emission, then \eqref{eq:5} gives an emission rate
from the up and down quarks greater than $10^{-10}$ eV, corresponding
to a distance of a few kilometers. That is, such high energy protons
would not travel more than a few kilometers before fragmenting and
losing their energy. The observation of these high energy protons is
inconsistent with this rapid energy loss. We conclude that the energy
of these cosmic protons must be below the threshold for Higgs
emission, $10^{20} \text{ eV} < E_{0}$. This provides our constraint
on positive $\delta$: $\delta < 10^{-18}$.

For negative $\delta$ the Higgs is superluminal and anomalous vacuum
emission processes by otherwise stable particles are forbidden by
energy and momentum conservation. However the modifications of the $W$
and $Z$ propagators lead to potentially observable effects on the
behavior of light hadrons and leptons. As in the subluminal case,
these effects are largest at high energies and the most stringent
constraints come from ultra-high-energy cosmic rays.

Weak decays of charged pions and muons are well described by the
tree-level exchange of the $W$ boson. The modification of the $W$
propagator in \eqref{eq:2} leads to a modification of the effective
interaction. Since $k^{2} \sim {\cal O}(m^{2}_{\pi}, m^{2}_{\mu}) \ll
M_{W}^{2}$ we may safely ignore powers of $k^{2}/M_{W}^{2}$. Also, the
light quark and leptonic weak currents are nearly conserved and
contributions from terms in the vector boson propagators involving
$k^{\mu}$ or $k^{\nu}$ are proportional to the tiny light quark and
lepton Yukawa couplings. With the neglect of such small effects the
effective Lagrangian for the charged-current weak interaction is
\begin{equation}
  \label{eq:3}
  {\cal L}_{wk} = \frac{G_{F}}{\sqrt{2}} J^{\mu}J^{\nu} 
  \left\{g_{\mu\nu} + n_{\mu}n_{\nu}\frac{M^{2}_{W}}{(k\cdot n)^{2} +
  E_{1}^{2}}\right\}
\end{equation}
where all reference to the Lorentz violating parameter $\delta$ has
been subsumed into the energy $E_{1}^{2} \equiv -M^{2}_{W}/\delta >
0$. The neutral current weak interaction is similarly modified.

The conventional $g^{\mu\nu}$ term in the interaction involves the
square of the weak current's Lorentz invariant length and is
proportional to the square of the small meson masses. The Lorentz
violating correction, however, involves the time component of this
current which grows with energy. Consequently, at high energies this
second term becomes much larger than the first and the decay rates of
light mesons are greatly enhanced. Notably, the decay is not helicity
suppressed at high energy and charged pions decay more often to
electrons than muons. An elementary calculation yields a charged pion
decay rate to lepton $l=e,\mu$ and an associated neutrino
\begin{equation}
  \label{eq:6}
  \Gamma = \frac{F_{\pi}^{2}}{6\pi E}
  \left(
    \frac{G_{F}M_{W}^{2}}{\sqrt{2}}
  \right)^{2}
  \left(
    \frac{E^{2}}{E^{2}+E_{1}^{2}}
  \right)^{2}
  K_l
\end{equation}
where the pion decay constant $F_{\pi}\sim 130 \text{ MeV}$, the weak
coupling $G_{F}M_{W}^{2}/\sqrt{2} = g_{\text{wk}}^{2}/8 \sim 1/20$,
and $K_{l} = (1-m_{l}^{2}/m_{\pi}^{2})^{2}(1+2 m_{l}^{2}/m_{\pi}^{2})$
is close to one for the electron and $0.4$ for the muon. A similar
expression holds for the rate of muon decay to
$e^{-}\bar{\nu}_{e}\nu_{\mu}$ (neglecting the electron and neutrino
masses and higher order weak effects):
\begin{equation}
  \label{eq:7}
  \Gamma = \frac{m_{\mu}^{2}}{240\pi^{3} E}
  \left(
    \frac{G_{F}M_{W}^{2}}{\sqrt{2}}
  \right)^{2}
  \left(
    \frac{E}{E_{1}} 
  \right)^{4}g(E/E_{1})
\end{equation}
where $g(x) = 10 [3\arctan(x)/x - 3 + \ln (1+x^{2})]/x^{4}$ approaches
one for small argument.

In a Lorentz invariant world high energy protons impinging on the
atmosphere produce a hadronic air shower which may be caricatured as
follows. At the highest energies, near $10^{20}$ eV, hundreds of
mesons are usually produced in the initial
collision\cite{Stanev:2003}; at these extreme energies the resulting
mesons have a greatly dilated lifetime, and charged mesons collide
with nuclei in the atmosphere long before they decay. These secondary
collisions degrade the energy per hadron and develop the cascade. With
the energy degraded the meson lifetime is decreased: some charged
mesons in the shower decay above the Earth's surface producing
long-lived muons that reach the ground. The neutral pions decay
rapidly producing high energy photons. The result is a long cascade
(whose length increases with the energy of the primary) containing
many muons.

An enhanced decay rate for mesons would radically and visibly alter
the morphology of ultra-high-energy cosmic ray proton air showers.  At
sufficiently high energy even a very small $\delta$ would lead to
significant Lorentz violation. The charged pion lifetime would be
short and these particles would decay before they interact, resulting
in high energy charged leptons (about 2/3 electrons and 1/3 muons)
accompanied by high energy neutrinos. A small fraction of the mesons
produced in the initial collision are kaons, but they would have a
similarly short lifetime, decaying to muons, electrons and neutrinos.
If the muon lifetime is also short, high energy muons would decay
quickly and not reach the ground. Finally the neutral pion branching
fraction to neutrinos would be large, resulting in fewer high energy
photons in the shower. The result is a shallow shower with a rapid
flow of the energy of the initial proton into electrons and neutrinos:
there would no longer be a hadronic cascade.

Pions and muons produced in the initial collision of a primary proton
with energy near $10^{20}$ eV have energies of order or greater than
$10^{17}$ eV. For this energy a value of $\delta = -2\times 10^{-14}$
gives a charged pion decay length of about 5 meters, a muon decay
length of about 5 kilometers, and a neutral pion branching fraction to
neutrinos of about 85\%. A proton-initiated air shower with these
characteristics does not resemble actual ultra-high-energy cosmic ray
showers, even in caricature. The observation of conventional high
energy showers with their long development length and significant muon
component rules out such a large value of $\vert \delta \vert$.

A more detailed evaluation of the morphology of ultra-high-energy
cosmic ray air showers and the spectrum of charged leptons and
neutrinos produced will certainly strengthen the constraints presented
here. Similar arguments can be used to constrain other forms of
Lorentz violation as well, including cases that do not maintain
rotational symmetry.

In a Lorentz invariant world physics depends only on Lorentz covariant
quantitites: invariants depend only on lengths of four-vectors; rates
fall at high energies according to time dilation.  But the effects of
small departures from Lorentz symmetry generally grow with energy, and
the extreme particle energies accessible in cosmic rays provide the
ideal environment to seek violations of Lorentz invariance. These
experiments may yet provide the first evidence for departures from
relativistic invariance.

\begin{acknowledgments} 
  We thank Shelly Glashow, Ed Kearns, and Martin Schmaltz for useful
  discussions.  This work was supported by the U.S.\ Department of
  Energy's Office of Science.
\end{acknowledgments}

\bibliography{higgs_speed}
\bibliographystyle{utphys}

\end{document}